\documentstyle[11pt]{article}
\begin{document}

\title{Pair production of charged Higgs bosons in gluon-gluon collisions}
\author{{
Jiang Yi$^{b}$, Han Liang$^{b}$, Ma Wen-Gan$^{a,b}$,
Yu Zeng-Hui$^{b}$ and Han Meng$^{b}$ }\\
{\small $^{a}$CCAST (World Laboratory), P.O.Box 8730, Beijing 100080,} \\
{\small People's Republic of China} \\
{\small $^{b}$Modern Physics Department, University of Science and}\\
{\small Technology of China, Anhui 230027,}  \\
{\small People's Republic of China. \footnote{Mailing address} } }
\date{}
\maketitle

\begin{center}\begin{minipage}{100mm}
\vskip 10mm
\begin{center} {\bf Abstract}\end{center}
\baselineskip 0.3in
{We studied the pair production rate of charged Higgs bosons via
gluon-gluon collisions in the general two-Higgs-doublet model.
The analysis of its production are made with possible parameter
values of Higgs sectors in this model. We find that the production
cross sections may be 4 to 41 fermi-bar in the future LHC pp collider
with these parameters. The calculation also shows that this pair
production process is not sensitive to the neutral Higgs masses,
therefore its measurment could not give more imformations about the
neutral Higgs bosons.}

\vskip 10mm
{PACS number(s): 14.80.Cp, 12.15.Lk, 12.60.Cn}
\end{minipage}
\end{center}

\baselineskip=0.36in

\newpage
\noindent
{\Large{\bf 1.Introduction}}
\vskip 5mm

\begin{large}
\baselineskip 0.35in
In the sense of Minimal Standard Model(MSM) \cite{s1}\cite{s2}, a neutral
Higgs boson is predicted without giving its mass value. Therefore
the search for Higgs boson is undoubtedly one of
the important issues in testing the electroweak theory.
Untill now, the top quark mass is given as \cite{s3}: $m_t = 176\pm
10^{+8}_{-10} GeV$. The absence of Higgs-boson
signal events leads to a lower bound $M_{H^{0}} > 63.5 GeV$ at $95\%$
C.L. obtained \cite{s4}. In order to investigate the sectors of the
theory about which we know very little, the future multi-TeV hadron
colliders such as the CERN Large Hadron Collider (LHC) and Next Linear
Collider(NLC) are elaborately designed to devote to the precise
characters of Higgs sector and top quark. Besides providing
the first evidence of Higgs mechanism, the multi-TeV
colliders offer the possibility of carrying on the research of
other unknown particles in the hitherto unexplored mass range.

So far the extension of the MSM which includes a second Higgs doublet
is far from being rules out experimentally. One strong motivation for
extending Higgs sector into two-Higgs-doublet model (THDM), comes from
that it has a common feature of many extended models of the MSM, such as
Minimal Supersymmetric Standard Model(MSSM). The MSSM model requires
two SU(2) doublets to give masses to up and down quarks\cite{s5}.
It is therefore of interest to determine whether THDM or MSSM is
compatible with experiments, and what further extensions should be
if it is not.

In future hadron colliders, several mechanisms can produce
Higgs boson. Among them the Higgs pair productions via gluon-gluon
collision, quark-antiquark annihilation or others, play a significant
role in studing Higgs sectors. The on-mass shell charged Higgs boson pair
production with their subsequent cascade decays will provide a good
opportunity of studying the two Higgs doublet model and a copious
source of charged Higgs bosons. Of cause, the Higgs pair production
has less rate comparing with the single production mechanism,
but the Higgs pair production via gg and $q\bar{q}$ annihilation involves
the trilinear and quartic couplings of the physical Higgs bosons. So
the measurement of these couplings may be a good test of our theory.
At the mean time, the pair production of Higgs boson in hadron
collisions has a better signal-to-background ratio than one Higgs
production process. Therefore it is worthwhile to study the precise features
of the Higgs pair production processes at LHC, NLC and other future colliders
which are designed operating in $\gamma \gamma$,$e^{+} e^{-}$, $pp$ and
$p \bar{p}$ collision modes. Especially when charged Higgs boson mass is
not very heavy, its pair production will be more significant.
Recently T. Plehn, M. Spira and P.M. Zerwas provided the calculations
of pair production of neutral Higgs particles in gluon-gluon collisions
in MSSM\cite{s6} and charged Higgs boson pair production at NLC in
$\gamma \gamma$ modes are also investigated in references \cite{s7}.

In this paper we represent the calculation of the charged Higgs boson pair
production via gluon-gluon collisions at LHC in the frame of general THDM.
The paper is organized as follows: The analytical evaluations
are given in section 2. In section 3 there are numeriacal results,
discussion and a short summary. Finally the explicit expressions of
the relevant form factors are collected in Appendix.

\vskip 5mm
\noindent
{\Large{\bf 2.Calculation}}
\vskip 5mm
The general THDM contains six free parameters. Generally the four
mass parameters, i.e. the parameters for two scalar neutral Higgs
boson masses $m_{h^{0}}$ and
$m_{H^{0}}$, one neutral pseudoscalar Higgs mass $m_{A^{0}}$, one pair
charged Higgs boson masses $m_{H^{\pm}}$ and the additional mixing
angle $\alpha$ and $\beta$, are chosen as free parameters.
The parameter $\alpha$ is defined as the mixing angle of the scalar
Higgs bosons $h^{0}$ and $H^{0}$. The angle $\beta$ is a mixing
parameter of charged and pseudoscalar Higgs bosons and is related
directly to the vacuum expectation values of Higgs doublets
(i.e. $\tan\beta=v_{2}/v_{1}$). In order to demonstrate the numerical
features of the charged Higgs boson pair production, we specify
$\alpha=\beta$ in our numerical calculation for simplicity.

The generic Feynman diagrams contributing to the subprocess
$gg \rightarrow H^{+}H^{-}$ which are involved in frame of
THDM  are depicted in figure 1. The relevant Feynman rules
can be found in references\cite{s5}. Since we neglect the
couplings between light-quarks and Higgs bosons, there are
only the following three kinds of production mechanisms: One
way is by exchanging a $\gamma$ or a $Z^0$ boson to produce a charged
Higgs boson pair(see Fig.1(a,b)), we call it as
$\gamma-Z^0$ s-channel. The $\gamma$ exchange s-channel diagrams have
no contribution and only the axis-vector part of the coupling between
$Z^0$ and the heavy quark in $Z^0$ exchange diagram, contributes
to the cross section. The second mechanism is through producing the
virtual neutral Higgs $h^0$ or $H_{0}$ bosons which is coupled to
gluons by the usual heavy-quark triangle, and the charged Higgs boson
pair is appeared by the subsequent decay of the virtual Higgs bosons
(see Fig.1(c,d)), which are called the $h^0-H^0$ s-channel. Another
mechanism is named as heavy-quark box diagrams(see Fig.1($e \sim j$)).
The box-diagrams of exchanging the two initial gluons are not
plotted in Fig.1. All the box diagrams can be divided into
two groups. They are corresponding to t-channel and u-channel
groups respectively. In this work, we used n-dimensional
regularization scheme in the one-loop diagram calculation.

We denote $\theta$ as the scattering angle between the one of
the gluons and final $H^{+}$ boson. Then in the system of
center-of-mass (CMS) we can express all the four-momenta of
initial and final particles by means of the total energy
$\sqrt{\hat{s}}$, the scattering angle $\theta$ and the
velocity of final charged Higgs bosons. In the CMS of the
initial two gluons, the four-momenta of the final charged
Higgs bosons read in term of following forms
$$ p_1=(\frac{\sqrt{\hat{s}}}{2},\frac{\sqrt{\hat{s}}}{2}
     \beta_{H^{\pm}}\sin{\theta}, 0,\frac{\sqrt{\hat{s}}}{2}
     \beta_{H^{\pm}}\cos{\theta}), $$
$$ p_2=(\frac{\sqrt{\hat{s}}}{2},\frac{-\sqrt{\hat{s}}}{2}
     \beta_{H^{\pm}}\sin{\theta}, 0,\frac{-\sqrt{\hat{s}}}{2}
     \beta_{H^{\pm}}\cos{\theta}). \eqno(1) $$
where the velocity of charged Higgs bosons in the CMS,
$\beta_{H^{\pm}}$, can be represented as
$$
\beta_{H^{\pm}}=\sqrt{1-4 m_{H^{\pm}}^2/\hat{s}}.  \eqno(2)
$$
$p_3$ and $p_4$ are the four-momenta of the initial gluons and
expressed as
$$
p_3=(\frac{\sqrt{\hat{s}}}{2},0,0,\frac{\sqrt{\hat{s}}}{2}) ,
$$
$$
p_4=(\frac{\sqrt{\hat{s}}}{2},0,0,\frac{-\sqrt{\hat{s}}}{2}). \eqno(3)
$$
The corresponding matrix element for all the diagrams in figure 1
is written as
\begin{eqnarray*}
M  &=& M^{\hat{s}} + M^{box(\hat{t})} + M^{box(\hat{u})} \\
   &=& \frac{-1}{16 \pi^2} T^a T^b g_{s}^{2} \epsilon_{\mu}(p_3)
           \epsilon_{\nu}(p_4) \\
    & &   (f_{1} g^{\mu \nu} + f_{2} p_1^{\mu} p_1^{\nu} +
          f_{3} p_1^{\mu} p_2^{\nu} +
          f_{4} p_2^{\mu} p_1^{\nu} + f_{5} p_2^{\mu} p_2^{\nu} + \\
    & &   f_{6} \epsilon^{\mu \nu \alpha \beta} p_{1 \alpha} p_{2 \beta} +
          f_{7} \epsilon^{\mu \nu \alpha \beta} p_{1 \alpha} p_{3 \beta} +
          f_{8} \epsilon^{\mu \nu \alpha \beta} p_{2 \alpha} p_{3 \beta} + \\
    & &   f_{9} \epsilon^{\nu \alpha \beta \gamma} p_{1 \alpha}
                    p_{2 \beta} p_{3 \gamma} p_1^{\mu} +
          f_{10} \epsilon^{\mu \alpha \beta \gamma} p_{1 \alpha}
                    p_{2 \beta} p_{3 \gamma} p_1^{\nu} +
\end{eqnarray*}
$$
   f_{11} \epsilon^{\nu \alpha \beta \gamma} p_{1,\alpha}
   p_{2,\beta} p_{3,\gamma} p_2^{\mu} +
   f_{12} \epsilon^{\mu \alpha \beta \gamma} p_{1,\alpha}
   p_{2,\beta} p_{3,\gamma} p_2^{\nu} ).  \eqno(4)
$$

The upper indexes $\hat{s}$, box($\hat{t}$) and box($\hat{u}$)
represent the amplitudes corresponding to the s-channel diagrams,
including all the $\gamma-Z^0$ and $h^0-H^0$ diagrams, t-channal
and u-channal box diagrams in Fig.1 respectively.
The $g_s$ is the strong coupling constant and $T^a(a=1 \sim 8)$ are
the $SU(3)_C$ generators introduced be Gell-Mann\cite{s8}.
The variables $\hat{s}$, $\hat{t}$ and $\hat{u}$ are usual Mandelstam
variables in the gluon-gluon center-of mass system with the definitions

$$ \hat{s}=(p_1+p_2)^2=(p_3+p_4)^2, $$
$$ \hat{t}=(p_1-p_3)^2=(p_2-p_4)^2, $$
$$ \hat{u}=(p_1-p_4)^2=(p_2-p_3)^2. \eqno(5)  $$

We collect the explicit expressions of the form factors
$f_{i}(i=1 \sim 12)$ which appear in equation (4) in Appendix.
The total cross section for subprocess $gg \rightarrow H^{+}H^{-}$
can finally be written in the form
$$
\hat{\sigma}(\hat{s})=\frac{1}{16 \pi \hat{s}^{2}}
\int_{\hat{t}^{-}}^ {\hat{t}^{+}} d\hat{t} \bar{\vert M \vert}^2. \eqno(6)
$$
where $\bar{\vert M \vert}^2$ is the initial spin and color averaged matrix
element squared and $\hat{t}^{\pm}=(m_{H^{\pm}}^2-\frac{1}{2}\hat{s})
\pm \frac{1}{2}\hat{s} \beta_{H^{\pm}}$. The color factor including in
$\bar{\vert M \vert}^2$ is equal to 1/32. The total cross section for
the charged Higgs pair production through gluon-gluon fusion in
proton-proton collisions can be obtained by integrating the
$\hat{\sigma}$ over the gluon luminosity.
$$
\sigma(pp \rightarrow gg \rightarrow H^{+}H^{-}+X)=
    \int_{4 m_{H^{\pm}}^2/s }^{1}
d\tau \frac{dL_{gg}}{d\tau} \hat{\sigma}(gg \rightarrow
H^{+}H^{-} \hskip 3mm at \hskip 3mm \hat{s}=\tau s). \eqno(7)
$$
Here we introduced a variable $\tau=x_{1}x_{2}$ and $x_1$ and $x_2$ are
defined as
$$
x_{1}=\frac{p_3} {P_{1}}, ~~x_{2}=\frac{p_4} {P_{2}},
$$
where $P_1$ and $P_2$ are the four momentum of both incident protons
respectively. $\sqrt{s}$ and $\sqrt{\hat{s}}$ are the proton-proton
and gluon-gluon center-of-mass energies respectively, which can be
expressed as
$$
s=(P_1+P_2)^2,~~ \hat{s}=(p_3+p_4)^2=x_1 x_2 s=\tau s. \eqno(8)
$$
The quantity of the gluon luminosity $\frac{dL_{gg}} {d\tau}$ can be
calculated by using the following equation.
$$
\frac{dL_{gg}}{d\tau}=2 \int_{\tau}^{1}
 \frac{dx_1}{x_1} \left[ F_{g}(x_1,\mu) F_{g}(\frac{\tau}{x_1},\mu) \right].
 \eqno(9)
$$
  The numerical resuts of the process $pp \rightarrow gg \rightarrow
H^{+}H^{-}$ are represented by adopting the MRS set G parton distribution
function $F_{g}(x)$ \cite{s9}.

\vskip 5mm
\noindent
{\Large{\bf 3. Numerical results and discussion}}
\vskip 5mm

In the numerical evolution we take the input parameters\cite{s10} as
$m_b=4.5~ GeV,~m_Z=91.1887~GeV,~m_W=80.2226~GeV$,  ~$G_F=1.166392\times
10^{-5}(GeV)^{-2}$ and $\alpha=\frac{1}{137.036}$. The top quark mass is
just fixed to be the central value $m_t=175~GeV$. We also used a simple
one-loop formula for the running strong coupling constant $\alpha_s$ as
$$
\alpha_s(\mu)=\frac{\alpha_{s}(m_Z)} {1+\frac{33-2 n_f} {6 \pi} \alpha_{s}
              (m_Z) \ln \left( \frac{\mu}{m_Z} \right) }. \eqno(10)
$$
where $\alpha_s(m_Z)=0.117$ and $n_f$ is the number of active flavors at
scale $\mu$. The ratio of the two vacuum expectation values $\tan{\beta}$
is chosen to be varied in our calculation. The coupling
strength of the charged Higgs boson with quarks is related
to scalar and pseudo-scalar parts, each part has factor $g_{+}$ or
$g_{-}$(The definitions of $g_{+}$ and $g_{-}$
can be found in Appendix). These two parts of Yukawa coupling yield
a crucial contribition alternatively depending on variate $\tan{\beta}$.
Because the present experimental data disfavor the small values of
$\tan{\beta}$ in general THDM. Therefore we limit its quantity in
the representative range of 1.5 to 30 in the calculation.

The cross sections of the subprocess $gg \rightarrow H^{+} H^{-}$
with $m_{H^{\pm}} = 150~GeV$ are depicted in Fig.2(1), on the conditions
of $m_{h^{0}} = 100~GeV$ and $m_{H^{0}} = 150~GeV$. Fig.2(2) shows also
the cross sections of this subprocess as the functions of CMS energy
of incoming gluons with the same neutral Higgs boson masses, but  
$m_{H^{\pm}} = 300~GeV$. In both figures, the full line curves
and dashed line curves are corresponding to the $\tan{\beta}=1.5$
and $\tan{\beta}=30$ respectively. These graphs show that
the cross section of the subprocess is approximately two times
larger when the ratio of vaccum expectation values is 30 than
that when $\tan{\beta}=1.5$ in the energy region less than 2 TeV.
In Fig.2(1) there is a peak around $\sqrt{\hat{s}} \sim 400~GeV$.
This peak comes from the $t \bar{t}$ threshold effects and enhanced
by the top quark loop diagrams.

The calculation shows the result of the cross section $\hat{\sigma}$
is not sensitive to the masses of the related two neutral Higgs bosons
$m_{h^{0}}$ and $m_{H^{0}}$, especially when $\sqrt{\hat{s}}$ is far beyond
the value regions of the two masses $m_{h^{0}}$ and $m_{H^{0}}$. From the
Feynman diagrams in Fig.1, we know that the diagrams, which are related
to the neutral Higgs bosons and the trilinear Higgs couplings, are only
appeared in the s-channel as shown in Fig.1(a,b), but their contribution
is negligible quantitatively. This comes from two reasons. Firstly,
the contribution is enhanced mainly by the Yukawa coupling which
is proportional to $m_t/m_W$. Since the $\gamma$ exchange s-channel
diagram has no contribution and $Z^0$ boson exchange s-channel diagram
does not include Yukawa coupling, the contribution from these kinds of
s-channel diagrams is rather small. Among the s-channel diagrams,
Fig.1(b) has only one Yukawa coupling, whereas each
box diagram in Fig.1($c \sim h$) has two such couplings, therefore
the cross section of the subprocess is mainly contributed by the
box diagrams. Secondly, when the energy of CMS system $\sqrt{\hat{s}}$
is not in the regions arround $m_{h^0}$ and $m_{H^0}$, the propagators
in s-channel diagrams will suppress their contributions to the cross
section. So it is clear that the measurement of this process could
not be expected to give constraint on the features of neutral Higgs
bosons and the trilinear Higgs couplings at the future pp colliders.

In figure 3 we show the cross section $\hat{\sigma}$ as the function of
$\tan{\beta}$ with $m_{H{\pm}}=150~GeV$, when $\sqrt{\hat{s}}=400~GeV$
and $\sqrt{\hat{s}}=800~GeV$ respectively. The cross section goes down
when $\tan{\beta}$ varies from 1.5 to 6, and raises up when $\tan{\beta}$
varies from 6 to 30. The cross sections are extremely small in the
region arround $\tan{\beta}=6$. Fig.4 is the plot of cross section
$\hat{\sigma}$ as the function of the charged Higgs boson mass with
$\sqrt{\hat{s}}=1~TeV$, $\tan{\beta}$ are 1.5 and 30 respectively.
There the cross section increases with increasing mass of the
charged Higgs boson $m_{H^{\pm}}$. In the region where
$m_{H^{\pm}} < 300~GeV$, the curves are flat, whereas they
raise up quickly when $m_{H^{\pm}} > 300~GeV$. That is because
when the charged Higgs boson pair production threshold approaches
the CMS energy $\sqrt{\hat{s}}=1~TeV$, the cross section will be
enhanced by the threshold effect of the charged Higgs boson pair
production.

In Fig.5(1) and Fig.5(2) we show the cross section of $pp \rightarrow
gg \rightarrow H^{+} H^{-} + X$ process with $m_{H^{\pm}}$ being $150~GeV$
and $300~GeV$ respectively. The charged Higgs boson pair production
rates are read to be about 14 to 41 fermi-bar at the
LHC energies when charged Higgs boson mass is 150 GeV and about
4.2 to 13.2 fermi-bar when $m_{H^{\pm}}=300~GeV$.

In conclusion, we have studied the pair production process of charged
Higgs boson via gluon-gluon fusion in pp collider at LHC. The numerical
analysis of its cross sections is carried out in the general two-Higgs-
doublet model. With the possible parameters, the cross sections at future
LHC collider can be 4 to 41 fermi-bar. These quantities are in the comparable
range with that of the neutral Higgs boson pair production in
gluon-gluon collisions as shown in Ref.\cite{s6}. The calculation shows also
that the production rate of charged Higgs boson pair is not sensitive
to the neutral Higgs boson masses and the trilinear Higgs couplings, but it
is strongly related to the couplings of the charged Higgs boson to fermions.

\par
This work was supported in part by the National Natural Science Foundation
of China and the National Committee of Science and Technology of China.
Part of this work was done when two of the authers, Ma Wen-Gan and
Yu Zeng-Hui, visited the University Vienna under the exchange
agreement(project number: IV.B.12).

\vskip 5mm
\noindent
{\Large{\bf Appendix}}
\vskip 5mm

We adopt the same definitions of one-loop integral functions as in
reference\cite{s11} and $d=4- \epsilon$. The integral functions are
defined as
$$
A_{0}(m)=-{\frac{(2\pi\mu)^{4-D}}{i\pi^2}}\int d^D q
{\frac{1}{[q^2-m^2]}},
$$
$$
\{B_0;B_{\mu};B_{\mu\nu}\}(p,m_1,m_2)=
{\frac{(2\pi\mu)^{4-D}}{i\pi^2}}\int d^D q
{\frac{\{1;q_{\mu};q_{\mu\nu}\}}{[q^2-m_1^2][(q+p)^2-m_2^2]}},
$$
$$
\{C_0;C_{\mu};C_{\mu\nu};C_{\mu\nu\rho}\}(p_1,p_2,m_1,m_2,m_3)=
$$
$$
-{\frac{(2\pi\mu)^{4-D}}{i\pi^2}}\int d^D q
{\frac{\{1;q_{\mu};q_{\mu\nu};q_{\mu\nu\rho}\}}
{[q^2-m_1^2][(q+p_1)^2-m_2^2][(q+p_1+p_2)^2-m_3^2]}},
$$
$$
\{D_0;D_{\mu};D_{\mu\nu};D_{\mu\nu\rho};D_{\mu\nu\rho\alpha}\}
(p_1,p_2,p_3,m_1,m_2,m_3,m_4)=
$$
$$
{\frac{(2\pi\mu)^{4-D}}{i\pi^2}}\int d^D q
{\frac{\{1;q_{\mu};q_{\mu\nu};q_{\mu\nu\rho};q_{\mu\nu\rho\alpha}\}}
{[q^2-m_1^2][(q+p_1)^2-m_2^2][(q+p_1+p_2)^2-m_3^2][(q+p_1+p_2+p_3)^2-m_4^2]}},
$$

The form factors involved in equation (4) are represented in terms of
the above integral functions explicitly as:

$$\begin{array}{lll}
f_1 &=& 2 m_t (g_{h^0 H^{\pm} H^{\mp}} g_{h^0tt} A_{h^0} +
        g_{H^0 H^{\pm} H^{\mp}} g_{H^0tt} A_{H^0})~
        d~~B_0 [p_1 + p_2, m_t, m_t] \\
    &-& 8 m_t (g_{h^0 H^{\pm} H^{\mp}} g_{h^0tt} A_{h^0} +
        g_{H^0 H^{\pm} H^{\mp}} g_{H^0tt} A_{H^0}) \cdot \\
    & & (d C_{24} + (p_4 \cdot p_3) C_0)[-p_3, p_1 + p_2, m_t, m_t, m_t]
         + i f_1^{-} g_{-}^2 + i f_1^{+} g_{+}^2,
\end{array}
\eqno{(A-1)}$$

where
$$ f_1^{-}=f_{1a}^{-}+f_{1b}^{-}+f_{1c}^{-}, \eqno{(A-2)}$$

\begin{eqnarray*}
f_{1a}^{-} &=& 2 ( 2 (m_b m_t C_0 - m_b^2 C_0 - m_{H^{\pm}}^2 C_{12}) +
      2 (p_1 \cdot p_3) (C_0 + C_{11} - C_{12}) - \\
    & &  2 (p_2 \cdot p_3) (C_0 + C_{11}) +
         2 (p_1 \cdot p_2) (C_{11} - C_{12}) - d C_{24})
         [p_1 - p_3, -p_4, m_b, m_t, m_t] \\
    &+& 2 (2 (p_1 \cdot p_3) (C_{11} - C_0 - C_{12}) -
         2 (p_1 \cdot p_2) C_{11} +
         2 (p_2 \cdot p_3) (C_0 + 3 C_{11} - 2 C_{12}) - \\
    & & d C_{24} + 2 m_b (m_t - m_b) C_0 - 4 m_{H^{\pm}}^2 C_{11} )
         [-p_2 + p_3, -p_3, m_b, m_t, m_t] \\
    &+& 4 (2 m_{H^{\pm}}^2 (D_{27} + 2 D_{311} - 2 D_{312} + 2 D_{313}) +
        4 m_t (m_t - m_b) D_{27} + \\
    & & 2 m_{H^{\pm}}^2 m_t^2 (D_{11} - D_{12} + D_{13}) +
        2 (p_1 \cdot p_2) (D_{27} + 2 D_{313} + m_t^2 D_{13}) +\\
    & & (p_1 \cdot p_3) (4 D_{312} - 4 D_{313} - 2 D_{27} +
         m_{H^{\pm}}^2 (D_{11} - D_{12} + 2 D_{21} + 2 D_{22} +\\
    & &   2 D_{23} - 4 D_{24} + 4 D_{25} - 4 D_{26}) + m_b m_t (2 D_{12} -
         D_0 - 2 D_{11} - 2 D_{13}) + \\
    & &  m_t^2 (D_0 + 2 D_{11}) ) +
        2 (p_1 \cdot p_2) (p_1 \cdot p_3) (D_{23} + D_{25} - D_{26}) +\\
    & & 2 (p_1 \cdot p_3)^2 (D_{24} - D_{22} - D_{23} - D_{25} + 2 D_{26}) +\\
    & & (p_2 \cdot p_3) (m_{H^{\pm}}^2 (D_{11}-D_{12}+2 D_{13}+2 D_{23}+
         2 D_{25} - 2 D_{26}) -\\
    & &  m_b m_t (D_0 + 2 D_{13}) + m_t^2 (D_0 + 2 D_{13}) ) +
         2 (p_1 \cdot p_2) (p_2 \cdot p_3) (D_{13} + D_{23}) +\\
    & &  2 (p_1 \cdot p_3) (p_2 \cdot p_3) (D_{26} - D_{23}) )
         [-p_1, p_1 - p_3, -p_4, m_t, m_b, m_t, m_t] \\
    &+& 4 (4 m_{H^{\pm}}^2 D_{311} + 2 m_{H^{\pm}}^4 (D_{11}+D_{21}+
         D_{24})-4 m_b m_t D_{27} +\\
    & & 4 m_t^2 D_{27} - 2 m_b m_t m_{H^{\pm}}^2 (D_0 + D_{11} + D_{12}) +
        2 m_{H^{\pm}}^2 m_t^2 (D_0 + 2 D_{11} + D_{12}) +\\
    & & 2 (p_1 \cdot p_2) (2 D_{312} + m_{H^{\pm}}^2 (D_{11} + D_{12} +D_{21}+
         D_{22} + 2 D_{24}) - m_b m_t (D_0 + D_{11} + D_{12}) + \\
    & &   m_t^2 (D_0 + D_{11} + 2 D_{12}) ) +
        2 (p_1 \cdot p_2)^2 (D_{12} + D_{22} + D_{24}) +
        (p_1 \cdot p_3) (2 D_{27} - 4 D_{312} + \\
    & & 4 D_{313} + m_{H^{\pm}}^2 (2 D_{13} - D_{11} - D_{12} - 2 D_{21} -
        2 D_{22} - 4 D_{24} + 4 D_{25} + 2 D_{26}) +\\
    & &  m_b m_t (D_0 + 2 D_{11} + 2 D_{12} - 2 D_{13}) +
         m_t^2 (4 D_{13} - D_0 - 2 D_{11} - 4 D_{12}) ) +\\
    & &  2 (p_1 \cdot p_2) (p_1 \cdot p_3) (D_{13} - D_{12} - 2 D_{22} -
           2 D_{24} + D_{25} + 2 D_{26}) +\\
    & &  2 (p_1 \cdot p_3)^2 (D_{22} + D_{23} + D_{24} - D_{25} - 2 D_{26}) + \\
    & &  (p_2 \cdot p_3) (m_{H^{\pm}}^2 (2 D_{25} - D_{11} - D_{12} - 4 D_{24}) +
          m_b m_t (D_0 + 4 D_{12} - 2 D_{13}) +\\
    & &   m_t^2 (2 D_{13} - D_0 - 4 D_{12}) ) +
         2 (p_1 \cdot p_2) (p_2 \cdot p_3) (D_{26} - D_{12} - 2 D_{22}) + \\
    & &  2 (p_1 \cdot p_3) (p_2 \cdot p_3) (D_{23} + 2 D_{22} - 3 D_{26}) )
        [-p_1, -p_2 + p_3, -p_3, m_t, m_b, m_t, m_t],
\hskip 9mm (A-3)
\end{eqnarray*}

$$ f_{1b}^{-}=f_{1a}^{-}~~(m_{b}\leftrightarrow m_{t}), \eqno{(A-4)}$$

\begin{eqnarray*}
f_{1c}^{-} &=& 2 d (B_0 [-p_3, m_t, m_t] + B_0 [-p_3, m_b, m_b] -
                 B_0 [-p_2, m_t, m_b]) \\
    &+& 4 (m_{H^{\pm}}^2 (C_{11} + C_{12}) +
        (p_1 \cdot p_2) (C_{11} + C_{12}) -
        (p_1 \cdot p_3) (2 C_0 + 3 C_{11}) - \\
    & & (p_2 \cdot p_3) (C_0 + C_{11} + 2 C_{12}) )
         [-p_1 + p_3, -p_2, m_b, m_t, m_b] \\
    &+& 4 (m_{H^{\pm}}^2 (C_{11} - C_0 - C_{12}) +
         (p_1 \cdot p_2) (C_{11} - C_0 - C_{12}) -
         (p_1 \cdot p_3) C_{11} + \\
    & & (p_2 \cdot p_3) (2 C_{12} - C_0 - 3 C_{11}) )
         [p_2 - p_3, -p_2, m_b, m_t, m_b] \\
    &+& 4 (2 (m_b^2 - 2 m_b m_t + m_t^2) D_{27} +
        2 m_{H^{\pm}}^2 m_t^2 (D_{12} - D_{11}) +
        2 (p_1 \cdot p_2) (D_{27} + m_t^2 D_{13}) + \\
    & & (p_1 \cdot p_3) (2 D_{27} + m_b^2 (D_{11} - D_{12}) +
         m_{H^{\pm}}^2 (D_{13} - 2 D_{21} - 2 D_{22} + 4 D_{24}) +\\
    & &  m_t^2 (D_0 - D_{11} - D_{12}) - m_b m_t D_0 ) +
        2 (p_1 \cdot p_2) (p_1 \cdot p_3) (D_{25} - D_{26}) +\\
    & & 2 (p_1 \cdot p_3)^2 (D_{22} - D_{24}) +
        (p_2 \cdot p_3) (m_{H^{\pm}}^2 (D_{11} - D_{12} + 2 D_{25} - 2 D_{26}) -\\
    & &  m_b^2 D_{13} - m_b m_t D_0 + m_t^2 (D_0 + D_{13}) ) +
        2 (p_1 \cdot p_3) (p_2 \cdot p_3) D_{26} -\\
    & & 2 (p_1 \cdot p_2) (p_2 \cdot p_3) (D_{13} + D_{23}) )
       [p_1, -p_1 + p_3, -p_2, m_t, m_b, m_t, m_b]  \\
    &+& 4 (2 (m_b^2 + m_{H^{\pm}}^2 + m_t^2) D_{27} - 4 m_b m_t D_{27} +
        m_b^2 m_{H^{\pm}}^2 (D_{11} + D_{12} - D_{13}) +\\
    & & m_{H^{\pm}}^4 (D_{13} - D_{11} - D_{12} - 2 D_{21} - 2 D_{24} + 2 D_{25}) +\\
    & & m_{H^{\pm}}^2 m_t^2 (D_{13} - 3 D_{11} - D_{12}) +
        (p_1 \cdot p_2) (4 D_{27} + m_b^2 (D_{11} + D_{12} - D_{13}) +\\
    & &  m_{H^{\pm}}^2 (D_{13} - D_{11} - D_{12} - 2 D_{21} -
                 2 D_{22} - 2 D_{23} - 4 D_{24} + 4 D_{25} + 4 D_{26}) +\\
    & & m_t^2 (3 D_{13} - D_{11} - 3 D_{12}) ) +
         2 (p_1 \cdot p_2)^2 (D_{25} - D_{22} - D_{23} - D_{24} + 2 D_{26}) +\\
    & & (p_1 \cdot p_3) (m_t^2 (D_0 + D_{11} + 3 D_{12}) - m_b m_t D_0 -
                            m_b^2 (D_{11} + D_{12}) +\\
    & &  m_{H^{\pm}}^2 (2 D_{11} + 2 D_{12} - D_{13} + 2 D_{21} +
         2 D_{22} + 4 D_{24} - 2 D_{26} - 2 D_{27}) ) +\\
    & & 2 (p_1 \cdot p_2) (p_1 \cdot p_3) (D_{12} - D_{13} + 2 D_{22} +
                                              2 D_{24} - D_{25} - 2 D_{26}) -\\
    & & 2 (p_1 \cdot p_3)^2 (D_{12} + D_{22} + D_{24}) +
       (p_2 \cdot p_3) (m_b^2 (D_{13} - 2 D_{12}) - m_b m_t D_0 +\\
    & & m_{H^{\pm}}^2 (D_{11} + D_{12} + 4 D_{24} - 2 D_{25}) +
                            m_t^2 (D_0 + 2 D_{12} - D_{13}) ) +\\
    & & 2 (p_1 \cdot p_2) (p_2 \cdot p_3) (D_{23} + 2 D_{22} - 3 D_{26}) + \\
    & & 2 (p_1 \cdot p_3) (p_2 \cdot p_3) (D_{26} - D_{12} - 2 D_{22}) )
        [p_1, p_2 - p_3, -p_2, m_t, m_b, m_t, m_b],
        \hskip 12mm (A-5)
\end{eqnarray*}

$$ f_1^{+} = f_1^{-} [m_b \rightarrow - m_b].  \eqno{(A-6)} $$

$$\begin{array}{lll}
f_2 &=& 8 m_t (g_{h^0 H^{\pm} H^{\mp}} g_{h^0tt} A_{h^0} +
         g_{H^0 H^{\pm}H^{\mp}} g_{H^0tt} A_{H^0}) \cdot \\
    & & (C_{11} + 2 C_{21} + 4 C_{22} - 4 C_{23}) [-p_3, p_1 + p_2, m_t, m_t, m_t] +
          i f_2^{-} g_{-}^2 + i f_2^{+} g_{+}^2,
\end{array}
\eqno{(A-7)}$$

where
$$ f_2^{-}=f_{2a}^{-}+f_{2b}^{-}+f_{2c}^{-},\eqno{(A-8)}$$
$$\begin{array}{lll}
f_{2a}^{-} &=& 4 (C_0 + C_{11} - C_{12}) [p_1 - p_3, -p_4, m_b, m_t, m_t]\\
  &+&  4 (C_0 + C_{11} - C_{12}) [-p_2 + p_3, -p_3, m_b, m_t, m_t] \\
  &+& (8 (D_{27} + 4 D_{311} - 2 D_{312} + 2 D_{313}) +
      4 m_{H^{\pm}}^2 (D_{12} - D_{11} - 4 D_{21} - 2 D_{22} + 6 D_{24} -\\
  & & 4 D_{25} + 2 D_{26} - 4 D_{31} + 8 D_{310} +
       8 D_{34} - 8 D_{35} - 4 D_{36} - 4 D_{37}) +\\
  & & 4 m_b m_t (D_0 + 4 D_{11} - 4 D_{12} + 4 D_{13} + 4 D_{21} -
      4 D_{24} + 4 D_{25}) + 4 m_t^2 (4 D_{12} - D_0 - 4 D_{11} - \\
  & & 4 D_{13} - 4 D_{21} + 4 D_{24} - 4 D_{25}) +
      8 (p_1 \cdot p_2) (D_{22} - D_{24} - D_{26} + 2 D_{310} -
      2 D_{35} - 2 D_{37}) + \\
  & & 8 (p_1 \cdot p_3) (D_{22} + D_{23} - D_{24} + D_{25} - 2 D_{26} -
      4 D_{310} - 2 D_{34} + 2 D_{35} + 2 D_{36} + 2 D_{37}) +\\
  & & 8 (p_2 \cdot p_3) (D_{24} - D_{22} - D_{23} - D_{25} + 2 D_{26}) )
      [-p_1, p_1 - p_3, -p_4, m_t, m_b, m_t, m_t]\\
  &+& 4 (2 (D_{27} + 4 D_{311} - 2 D_{312} + 2 D_{313}) +
       m_{H^{\pm}}^2 (D_{12} - D_{11} - 4 D_{21} - 2 D_{22} +\\
  & &  6 D_{24} - 4 D_{25} + 2 D_{26} - 4 D_{31} + 4 D_{34} - 4 D_{35}) -
       m_t (m_t - m_b) (D_0 + 4 D_{11} - 4 D_{12} +\\
  & &  4 D_{13} + 4 D_{21} - 4 D_{24} + 4 D_{25}) + 2 (p_1 \cdot p_2)
  (D_{22} - D_{24} - D_{26} - 2 D_{310} - 2 D_{34} + 2 D_{36}) +\\
  & &   2 (p_1 \cdot p_3) (D_{24} - D_{22} - D_{23} - D_{25} +
        2 D_{26} + 4 D_{310} + 2 D_{34} - 2 D_{35} - 2 D_{36} - 2 D_{37}) +\\
  & &   2 (p_2 \cdot p_3) (D_{22} + D_{23} - D_{24} + D_{25} - 2 D_{26}) )
       [-p_1, -p_2 + p_3, -p_3, m_t, m_b, m_t, m_t],
\end{array}
\eqno{(A-9)}$$

$$ f_{2b}^{-}=f_{2a}^{-}~~(m_{b}\leftrightarrow m_{t}),\eqno{(A-10)} $$
$$\begin{array}{lll}
f_{2c}^{-} &=& - 4 C_{11} [-p_1 + p_3, -p_2, m_b, m_t, m_b] -
            4 C_{11} [p_2 - p_3, -p_2, m_b, m_t, m_b] \\
  &+& 4 (m_b^2 (D_{12} - D_{11} - 2 D_{21} + 2 D_{24}) +
       m_t^2 (2 D_{24} - D_0 - 3 D_{11} + 3 D_{12} - 2 D_{21}) +\\
  & &   m_b m_t (D_0 + 4 D_{11} - 4 D_{12} + 4 D_{21} - 4 D_{24}) +
       m_{H^{\pm}}^2 (2 D_{26} - D_{13} - 2 D_{25}) - 2 D_{27} +\\
  & &   2 (p_1 \cdot p_2) (D_{22} - D_{24}) +
        2 (p_1 \cdot p_3) (D_{24} - D_{22}) +\\
  & &      2 (p_2 \cdot p_3) (D_{24} - D_{22} - D_{26}) )
       [p_1, -p_1 + p_3, -p_2, m_t, m_b, m_t, m_b] \\
  &+& 4 (m_b^2 (D_{12} - D_{11} - 2 D_{21} + 2 D_{24}) - 2 D_{27} +
        m_{H^{\pm}}^2 (2 D_{12} - D_{13} + 4 D_{24} - 2 D_{25}) +\\
  & &   m_b m_t (D_0 + 4 D_{11} - 4 D_{12} + 4 D_{21} - 4 D_{24}) +
       m_t^2 (3 D_{12} - D_0 - 3 D_{11} - 2 D_{21} + 2 D_{24}) +\\
  & &   2 (p_1 \cdot p_2) (D_{12} + D_{22} + D_{24} - D_{26}) -
        2 (p_1 \cdot p_3) (D_{12} + D_{22} + D_{24}) +\\
  & &   2 (p_2 \cdot p_3) (D_{26} - D_{12} - D_{22} - D_{24}) )
       [p_1, p_2 - p_3, -p_2, m_t, m_b, m_t, m_b],
\end{array}
\eqno{(A-11)}$$
$$ f_2^{+} = f_2^{-} [m_b \rightarrow - m_b].  \eqno(A-12) $$

$$\begin{array}{lll}
f_3 &=& 8 m_t (g_{h^0 H^{\pm} H^{\mp}} g_{h^0tt} A_{h^0} +
         g_{H^0 H^{\pm} H^{\mp}} g_{H^0tt} A_{H^0}) \cdot \\
    & & (C_{11} + 2 C_{21} + 4 C_{22} - 4 C_{23}) [-p_3, p_1 + p_2, m_t, m_t, m_t] +
         i f_3^{-} g_{-}^2 + i f_3^{+} g_{+}^2,

\end{array}
\eqno{(A-13)}$$

where
$$ f_3^{-}=f_{3a}^{-}+f_{3b}^{-}+f_{3c}^{-},\eqno{(A-14)}$$
$$\begin{array}{lll}
f_{3a}^{-} &=& 4 (C_0 - C_{12}) [-p_2 + p_3, -p_3, m_b, m_t, m_t] \\
  &+& 4 (2 C_{12} - 3 C_0 - 4 C_{11} - 2 C_{21} + 2 C_{23})
      [p_1 - p_3, -p_4, m_b, m_t, m_t]\\
  &+& 4 (4 (D_{27} + D_{312}) +
      m_{H^{\pm}}^2 (D_{12} - D_{11} - 4 D_{21} - 2 D_{22} - 4 D_{23} + 6 D_{24} -\\
  & & 8 D_{25} + 6 D_{26} - 8 D_{310} - 4 D_{32} - 4 D_{34} +
      8 D_{36} + 8 D_{38} - 4 D_{39}) +\\
  & & m_b m_t (D_{0} + 4 D_{11} - 4 D_{12} + 4 D_{13} - 4 D_{22} + 4 D_{24} +
      4 D_{26}) - m_t^2 (D_{0} + 4 D_{11} - \\
  & & 4 D_{12} + 4 D_{13} - 4 D_{22} + 4 D_{24} + 4 D_{26}) + 2 (p_1 \cdot p_2)
      (D_{22} - 2 D_{23} - D_{24} - 2 D_{25} + \\
  & & D_{26} - 2 D_{310} + 2 D_{38} - 2 D_{39}) + 4 (p_1 \cdot p_3) (D_{22} +
      D_{23} - D_{24} + D_{25} - \\
  & & 2 D_{26} + D_{310} + D_{32} - D_{36} - 2 D_{38} + D_{39}) )
      [-p_1, p_1 - p_3, -p_4, m_t, m_b, m_t, m_t]\\
  &+& 4 (2 (2 D_{313} - D_{27}) +
      m_{H^{\pm}}^2 (2 D_{26} - D_{11} - D_{12} - 4 D_{25} - 4 D_{35}) -\\
  & &  m_t (m_t - m_b) (D_0 + 4 D_{13} + 4 D_{25}) -
       2 (p_1 \cdot p_2) (D_{12} + D_{26} + 2 D_{310}) +\\
  & &  2 (p_1 \cdot p_3) (D_{12} - D_{13} - D_{23} + D_{26} +
       2 D_{310} - 2 D_{37}) +\\
  & &  2 (p_2 \cdot p_3) (D_{12} - D_{13} + D_{23} - D_{26}) )
       [-p_1, -p_2 + p_3, -p_3, m_t, m_b, m_t, m_t],
 \end{array}
\eqno{(A-15)}$$

$$ f_{3b}^{-}=f_{3a}^{-}~~(m_{b}\leftrightarrow m_{t}), \eqno{(A-16)}$$
$$\begin{array}{lll}
f_{3c}^{-} &=&  4 (C_0 - C_{12}) [p_2 - p_3, -p_2, m_b, m_t, m_b]
  + 4 (C_{12} + 2 C_0) [-p_1 + p_3, -p_2, m_b, m_t, m_b]\\
  &+& 4 (m_b^2 (D_{12} - D_{11} + 2 D_{22} - 2 D_{24} + 2 D_{25} - 2 D_{26}) +
       m_{H^{\pm}}^2 (2 D_{21} - D_{13} + \\
  & & 2 D_{22} - 4 D_{24}) - 4 D_{27} +
        m_b m_t (D_0 - 4 D_{22} + 4 D_{24} - 4 D_{25} + 4 D_{26}) +\\
  & &   m_t^2 (D_{11} - D_0 - D_{12} + 2 D_{22} - 2 D_{24} +
        2 D_{25} - 2 D_{26}) +\\
  & &   2 (p_1 \cdot p_2) (D_{22} - D_{24}) +
        2 (p_1 \cdot p_3) (D_{26} - D_{25} + 2 D_{24} - 2 D_{22}) +\\
  & &   2 (p_2 \cdot p_3) (D_{13} + D_{23} - D_{26}) )
        [p_1, -p_1 + p_3, -p_2, m_t, m_b, m_t, m_b] \\
  &+& 4 (m_{H^{\pm}}^2 (2 D_{13} + 2 D_{25} - D_{11} - D_{12}) - 2 D_{27} +
       m_b^2 (D_{13} + 2 D_{25}) + \\
  & &  m_b m_t (D_0 - 4 D_{13} -
   4 D_{25}) + m_t^2 (2 D_{25} - D_0 + 3 D_{13}) +
       2 (p_1 \cdot p_2) (D_{25} + D_{26} - \\
  & &     D_{12} + 2 D_{13} - D_{23}) +
     2 (p_1 \cdot p_3) (D_{12} - D_{13} - D_{25} - D_{26}) +\\
  & &   2 (p_2 \cdot p_3) (D_{12} - D_{13} + D_{23} - D_{26}) )
       [p_1, p_2 - p_3, -p_2, m_t, m_b, m_t, m_b],
\end{array}
\eqno{(A-17)}$$
$$ f_3^{+} = f_3^{-} [m_b \rightarrow - m_b].   \eqno(A-18) $$
$$\begin{array}{lll}
f_4 &=& 8 m_t (g_{h^0 H^{\pm} H^{\mp}} g_{h^0tt} A_{h^0} +
         g_{H^0 H^{\pm} H^{\mp}} g_{H^0tt} A_{H^0}) \cdot  \\
    & & (C_{11} + 2 C_{21} + 4 C_{22} - 4 C_{23}) [-p_3, p_1 + p_2, m_t, m_t, m_t] +
         i f_4^{-} g_{-}^2 + i f_4^{+} g_{+}^2,
 \end{array}
\eqno{(A-19)} $$

where
$$ f_4^{-}=f_{4a}^{-}+f_{4b}^{-}+f_{4c}^{-},\eqno{(A-20)}$$

$$\begin{array}{lll}
f_{4a}^{-} &=& 4 (C_0 - C_{12}) [p_1 - p_3, -p_4, m_b, m_t, m_t]\\
  &+& 4 (2 C_{12} - 3 C_0 - 4 C_{11} - 2 C_{21} + 2 C_{23})
      [-p_2 + p_3, -p_3, m_b, m_t, m_t]\\
  &+& 4 (2 (2 D_{313} - D_{27}) +
     m_{H^{\pm}}^2 (D_{12} - D_{11} - 2 D_{13} - 4 D_{25} + 2 D_{26} +\\
  & &   4 D_{310} - 4 D_{35} - 4 D_{37}) +
   m_b m_t (D_0 + 4 D_{13} + 4 D_{25}) - m_t^2 (D_0 + 4 D_{13} + 4 D_{25})-\\
  & &   2 (p_1 \cdot p_2) (D_{13} + D_{26} + 2 D_{37}) +
   2 (p_1 \cdot p_3) (D_{23} - D_{26} - 2 D_{310} + 2 D_{37}) +\\
  & & 2 (p_2 \cdot p_3) (D_{26} - D_{23}) )
      [-p_1, p_1 - p_3, -p_4, m_t, m_b, m_t, m_t]\\
  &+& 4 (4 (D_{27} + D_{312}) + m_{H^{\pm}}^2 (3 D_{12} - D_{11} -
      2 D_{13} - 4 D_{21} + 2 D_{22} + 2 D_{24} -\\
  & & 4 D_{25} - 2 D_{26} - 4 D_{310} - 4 D_{34} + 4 D_{36}) -
       m_t (m_t - m_b) (D_0 + 4 D_{11} - 4 D_{12} + \\
  & & 4 D_{13} - 4 D_{22} + 4 D_{24} + 4 D_{26}) +
  2 (p_1 \cdot p_2) (D_{12} - D_{13} + 3 D_{22} - 3 D_{24} - 3 D_{26} + \\
  & & 2 D_{32} - 2 D_{36} - 2 D_{38}) + 2 (p_1 \cdot p_3) (D_{13} -
      D_{12} - 2 D_{22} - 2 D_{23} + 2 D_{24} - 2 D_{25} + \\
  & & 4 D_{26} - 2 D_{310} - 2 D_{32} +
      2 D_{36} + 4 D_{38} - 2 D_{39}) + \\
  & &   2 (p_2 \cdot p_3) (D_{13} - D_{12}) )
      [-p_1, -p_2 + p_3, -p_3, m_t, m_b, m_t, m_t],

\end{array}
\eqno{(A-21)}$$

$$ f_{4b}^{-}=f_{4a}^{-}~~(m_{b}\leftrightarrow m_{t}), \eqno{(A-22)}$$

$$\begin{array}{lll}
f_{4c}^{-} &=&  4 (C_0 - C_{12}) [-p_1 + p_3, -p_2, m_b, m_t, m_b]
  + 4 (C_{12} + 2 C_0) [p_2 - p_3, -p_2, m_b, m_t, m_b]\\
  &+& 4 (m_{H^{\pm}}^2 (D_{12} - D_{11} + 2 D_{23}) - 2 D_{27} +
       m_b^2 (D_{13} + 2 D_{25}) + \\
  & & m_b m_t (D_0 - 4 D_{13} - 4 D_{25}) + m_t^2 (2 D_{25} - D_0 +
       3 D_{13}) + 2 (p_1 \cdot p_2) (D_{13} + D_{26}) + \\
  & & 2 (p_1 \cdot p_3) (D_{25} - D_{26}) -
       2 (p_2 \cdot p_3) (D_{23} + D_{26}) )
       [p_1, -p_1 + p_3, -p_2, m_t, m_b, m_t, m_b]\\
  &+& 4 (m_b^2 (D_{12} - D_{11} + 2 D_{22} - 2 D_{24} + 2 D_{25} - 2 D_{26}) +\\
  & &   m_{H^{\pm}}^2 (2 D_{21} - D_{13} + 2 D_{22} + 2 D_{23} - 2 D_{25} - 4 D_{26}) +\\
  & &   m_b m_t (D_0 - 4 D_{22} + 4 D_{24} - 4 D_{25} + 4 D_{26}) - 4 D_{27} +\\
  & &   m_t^2 (D_{11} - D_0 - D_{12} + 2 D_{22} - 2 D_{24} + 2 D_{25} - 2 D_{26}) +\\
  & &   2 (p_1 \cdot p_2) (D_{22} + D_{23} + D_{24} - D_{25} - 2 D_{26}) +
        2 (p_1 \cdot p_3) (D_{13} - 2 D_{24} + D_{25} + D_{26}) +\\
  & &   2 (p_2 \cdot p_3) (3 D_{26} - 2 D_{22} - D_{23}) )
        [p_1, p_2 - p_3, -p_2, m_t, m_b, m_t, m_b],
\end{array}
\eqno{(A-23)}$$
$$
f_4^{+} = f_4^{-} [m_b \rightarrow - m_b].   \eqno(A-24)
$$
$$\begin{array}{lll}
f_5 &=& 8 m_t (g_{h^0 H^{\pm} H^{\mp}} g_{h^0tt} A_{h^0} +
         g_{H^0 H^{\pm} H^{\mp}} g_{H^0tt} A_{H^0}) \cdot \\
    & & (C_{11} + 2 C_{21} + 4 C_{22} - 4 C_{23}) [-p_3, p_1 + p_2, m_t, m_t, m_t] +
         i f_5^{-} g_{-}^2 + i f_5^{+} g_{+}^2,

\end{array}
\eqno{(A-25)}$$

where
$$ f_5^{-}=f_{5a}^{-}+f_{5b}^{-}+f_{5c}^{-},\eqno{(A-26)}$$
$$\begin{array}{lll}
f_{5a}^{-} &=& 4 (C_0 + C_{11} + 2 C_{12} + 2 C_{23})[p_1 - p_3, -p_4, m_b, m_t, m_t]\\
  &+& 4 (C_0 + C_{11} + 2 C_{12} + 2 C_{23})[-p_2 + p_3, -p_3, m_b, m_t, m_t]\\
  &+& (4 m_{H^{\pm}}^2 (D_{12} - D_{11} - 2 D_{13} - 4 D_{23} - 4 D_{25} +
       2 D_{26} - 4 D_{310} + 4 D_{38} - 4 D_{39}) -\\
  & & 4 m_t (m_t - m_b) (D_0 + 4 D_{13} + 4 D_{26}) -
      8 (p_1 \cdot p_2) (D_{13} + 2 D_{23} + D_{26} + 2 D_{39}) +\\
  & & 16 (p_1 \cdot p_3) (D_{23} - D_{26} - D_{38} + D_{39}) )
      [-p_1, p_1 - p_3, -p_4, m_t, m_b, m_t, m_t]\\
  &+& 4 (m_{H^{\pm}}^2 (D_{12} - D_{11} - 2 D_{13} - 4 D_{25} - 2 D_{26} -
      4 D_{310}) - \\
  & &    m_t (m_t - m_b) (D_0 + 4 D_{13} + 4 D_{26}) -
     2 (p_1 \cdot p_2) (D_{13} + 3 D_{26} + 2 D_{38}) + \\
  & & 4 (p_1 \cdot p_3) (D_{26} - D_{23} + D_{38} - D_{39}) )
       [-p_1, -p_2 + p_3, -p_3, m_t, m_b, m_t, m_t],
 \end{array}
\eqno{(A-27)}$$

$$ f_{5b}^{-}=f_{5a}^{-}~~(m_{b}\leftrightarrow m_{t}), \eqno{(A-28)}$$
$$\begin{array}{lll}
f_{5c}^{-} &=& 4 (C_0 + C_{11}) [-p_1 + p_3, -p_2, m_b, m_t, m_b] +
      4 (C_0 + C_{11}) [p_2 - p_3, -p_2, m_b, m_t, m_b]\\
  &+& 4 (m_b^2 (D_{13} - 2 D_{23} + 2 D_{26}) +
       m_{H^{\pm}}^2 (D_{12} - D_{11} - 2 D_{25} + 2 D_{26}) +\\
  & &     m_b m_t (D_0 + 4 D_{23} - 4 D_{26}) +
       m_t^2 (2 D_{26} - D_0 - D_{13} - 2 D_{23}) +\\
  & &      2 (p_1 \cdot p_2) (D_{13} + D_{26}) -
   2 (p_1 \cdot p_3) (D_{13}+D_{26}) )
        [p_1, -p_1 + p_3, -p_2, m_t, m_b, m_t, m_b]  \\
  &+& 4 (m_{H^{\pm}}^2 (D_{12} - D_{11} - 2 D_{13} - 2 D_{25}) +
       m_b^2 (D_{13} - 2 D_{23} + 2 D_{26}) + \\
  & &  m_b m_t (D_0 + 4 D_{23} - 4 D_{26}) +
       m_t^2 (2 D_{26} - D_0 - D_{13} - 2 D_{23}) + \\
  & &  2 (p_1 \cdot p_3) (D_{13} + D_{26}) )
       [p_1, p_2 - p_3, -p_2, m_t, m_b, m_t, m_b],
\end{array}
\eqno{(A-29)}$$
$$
f_5^{+} = f_5^{-} [m_b \rightarrow - m_b].    \eqno(A-30)
$$

$$ f_6=f_{6a}+f_{6b}+f_{6c},\eqno{(A-31)}$$
$$\begin{array}{lll}
f_{6a} &=&
  g_{-} g_{+} (
      8 (C_{12} - C_{11}) [p_1 - p_3, -p_4, m_t, m_b, m_b]
  - 8 C_{11} [-p_2 + p_3, -p_3, m_b, m_t, m_t] \\
  &+& 16 ((p_1 \cdot p_3) (D_{12} - D_{11} - D_{13} - D_{21} - D_{22} -
       D_{23} + 2 D_{24} - 2 D_{25} + 2 D_{26} ) + \\
  & & (p_2 \cdot p_3) (D_{26} - D_{13} - D_{23} - D_{25}) - D_{27} )
      [-p_1, p_1 - p_3, -p_4, m_t, m_b, m_t, m_t] \\
  &+& 16 (m_{H^{\pm}}^2 (D_{12} + D_{22} + D_{24}) - 2 D_{27} +
      (p_1 \cdot p_2) (D_{12} + D_{22} + D_{24}) +\\
  & &    (p_1 \cdot p_3) (D_{26} - D_{11} - D_{21} - D_{22}) +\\
  & &  (p_2 \cdot p_3) (D_{26} - D_{12} -
       D_{22} - D_{24}) ) [-p_1, -p_2 + p_3, -p_3, m_t, m_b, m_t, m_t]),
 \end{array}
\eqno{(A-32)}$$
$$ f_{6b}=f_{6a}~~(m_{b}\leftrightarrow m_{t}), \eqno{(A-33)}$$

$$\begin{array}{lll}
f_{6c} &=&
      2 i g_s^2 g_{Z H^{\pm}H^{\mp}} g_{Ztt}^{a} A_{Z} \cdot
     \{ d B_0[-p_1-p_2,m_t,m_t] -
   4 C_{24}[p_4,-p_1-p_2,m_t,m_t,m_t] \} + \\
  & & g_{-} g_{+} (
   8 (C_{12} - C_{11}) [-p_1 + p_3, -p_2, m_b, m_t, m_b] \\
  &+& 8 C_0 + C_{11} - C_{12}) [p_2 - p_3, -p_2, m_b, m_t, m_b] \\
  &-& 16 D_{27} [p_1, -p_1 + p_3, -p_2, m_t, m_b, m_t, m_b] +
      8 (m_{H^{\pm}}^2 (D_{11} - D_{12} + D_{13} + 2 D_{21} - \\
  & &  2 D_{24} + 2 D_{25}) + m_b^2 (D_{12} - D_{11} - D_{13}) +
       m_t^2 (D_{11} - D_{12} + D_{13}) - 4 D_{27} + \\
  & & 2 (p_1 \cdot p_2) (D_{24} - D_{22} - D_{23} - D_{25} + 2 D_{26}) + \\
  & & 2 (p_1 \cdot p_3) (D_{22} - D_{24} - D_{26}) )
      [p_1, p_2 - p_3, -p_2, m_t, m_b, m_t, m_b] ).
\end{array}
\eqno{(A-34)}$$

$$ f_7=f_{7a}+f_{7b}+f_{7c},\eqno{(A-35)}$$
$$\begin{array}{lll}
f_{7a} &=&
   g_{-} g_{+} (
      8 (C_0 + C_{11} - C_{12}) [p_1 - p_3, -p_4, m_b, m_t, m_t] \\
  &+& 8 (C_0 + C_{11} - C_{12}) [-p_2 + p_3, -p_3, m_b, m_t, m_t] \\
  &+& 8 (m_{H^{\pm}}^2 (D_{11} - D_{12} + 2 D_{13} + 2 D_{21} +
  2 D_{22} + 4 D_{23} - 4 D_{24} + 6 D_{25} - 6 D_{26}) - \\
  & &  2 D_{27} - m_t^2 D_0 + 2 (p_1 \cdot p_2) (D_{11} - D_{12} + D_{13} +
      D_{21} + D_{22} + 2 D_{23} - 2 D_{24} + 3 D_{25} -\\
  & &    3 D_{26}) + 2 (p_4 \cdot p_3) (D_{24} - D_{22} - D_{23} -
     D_{25} + 2 D_{26}) ) [-p_1, p_1 - p_3, -p_4, m_t, m_b, m_t, m_t] \\
  &+& 8 (m_{H^{\pm}}^2 (D_{11} - D_{12} + 2 D_{13} + 2 D_{21} - 2 D_{22} +
       2 D_{25} + 2 D_{26}) - 2 D_{27} - m_t^2 D_0 + \\
  & &   2 (p_1 \cdot p_2) (D_{11} - D_{12} + D_{13} + D_{21} - D_{22} +
       D_{25} + D_{26}) + \\
  & &  2 (p_4 \cdot p_3) (D_{22} + D_{23} - D_{24} + D_{25} - 2 D_{26}) )
      [-p_1, -p_2 + p_3, -p_3, m_t, m_b, m_t, m_t] ).
 \end{array}
\eqno{(A-36)}$$

$$ f_{7b}=f_{7a}~~(m_{b}\leftrightarrow m_{t}) ,\eqno{(A-37)}$$

$$\begin{array}{lll}
f_{7c} &=&
     2 i g_s^2 g_{Z H^{\pm}H^{\mp}} g_{Ztt}^{a} A_{Z} \cdot
     \{ - d B_0[-p_1-p_2,m_t,m_t] + \\
  & & 4 [ (m_{H^{\pm}}^2 + p_1 \cdot p_2 - p_1 \cdot p_3 -
            p_2 \cdot p_3) C_{11} +
     d C_{24}][p_3,-p_1-p_2,m_t,m_t,m_t] \} \\
  &+& g_{-} g_{+} (
  - 8 (C_{11} + 2 C_0) [-p_1 + p_3, -p_2, m_b, m_t, m_b] -\\
  & &    8 (C_{11} + 2 C_0) [p_2 - p_3, -p_2, m_b, m_t, m_b] \\
  &+& 8 (m_{H^{\pm}}^2 (D_{13} - 2 D_{21} - 2 D_{22} + 4 D_{24}) +
       m_b^2 (D_{11} - D_{12}) +6 D_{27}+ \\
  & &      m_t^2 (D_0 - D_{11} + D_{12}) + 2 (p_1 \cdot p_2) (D_{25} - D_{26}) +
     2 (p_1 \cdot p_3) (D_{22} - D_{24}) - \\
  & &   2 (p_2 \cdot p_3) D_{13} )[p_1, -p_1 + p_3, -p_2, m_t, m_b, m_t, m_b] \\
  &+& 8 (m_{H^{\pm}}^2 (2 D_{12} - D_{13} - 2 D_{21} + 2 D_{24}) +
      m_t^2 (D_0 - D_{11} + D_{12}) +
        m_b^2 (D_{11} - D_{12}) +\\
  & &      6 D_{27} + 2 (p_1 \cdot p_2) (D_{12} - D_{13} + D_{22} - D_{24} +
       D_{25} - D_{26}) +
      2 (p_1 \cdot p_3) (D_{24} - \\
  & & D_{12} - D_{22}) + 2 (p_2 \cdot p_3)
  (D_{13} - D_{12}) ) [p_1, p_2 - p_3, -p_2, m_t, m_b, m_t, m_b] ).
\end{array}
\eqno{(A-38)}$$

$$ f_8=f_{8a}+f_{8b}+f_{8c},\eqno{(A-39)}$$
$$\begin{array}{lll}
f_{8a} &=&
   g_{-} g_{+} (
      8 (C_0 - C_{11}) [p_1 - p_3, -p_4, m_b, m_t, m_t] \\
  &+& 8 (C_0 - C_{11}) [-p_2 + p_3, -p_3, m_b, m_t, m_t] \\
  &+& 8 (m_{H^{\pm}}^2 (D_{12} - D_{11} + 4 D_{23} + 2 D_{25} - 2 D_{26}) -
      8 D_{27} - m_t^2 D_0 + \\
  & &  2 (p_1 \cdot p_2) (D_{25} + 2 D_{23} - D_{26}) +
       2 (p_1 \cdot p_3) (D_{24} - D_{22} - 2 D_{23} - D_{25} + \\
  & &     3 D_{26}) +  4 (p_2 \cdot p_3) (D_{26} - D_{23}) )
       [-p_1, p_1 - p_3, -p_4, m_t, m_b, m_t, m_t] \\
  &+& 8 (m_{H^{\pm}}^2 (D_{12} - D_{11} + 2 D_{22} + 2 D_{24}) -
      8 D_{27} - m_t^2 D_0 + \\
  & &  2 (p_1 \cdot p_2) (D_{22} + D_{24}) +
      2 (p_1 \cdot p_3) (D_{25} + D_{26} - D_{22} - D_{24}) + \\
  & &  4 (p_2 \cdot p_3) (D_{26} - D_{22}) )
      [-p_1, -p_2 + p_3, -p_3, m_t, m_b, m_t, m_t] ),
 \end{array}
\eqno{(A-40)}$$

$$ f_{8b}=f_{8a}~~(m_{b}\leftrightarrow m_{t}), \eqno{(A-41)}$$

$$\begin{array}{lll}
f_{8c} &=&
       2 i g_s^2 g_{Z H^{\pm}H^{\mp}} g_{Ztt}^{a} A_{Z} \cdot
        \{ d B_0[-p_1-p_2,m_t,m_t] + \\
  & & 4 [ (p_1 \cdot p_3 + p_2 \cdot p_3 - p_1 \cdot p_2 -
            m_{H^{\pm}}^2) C_{11} -
         d C_{24}][p_3,-p_1-p_2,m_t,m_t,m_t]\} \\
  &+& g_{-} g_{+} (
  - 8 (C_0 + C_{11}) [-p_1 + p_3, -p_2, m_b, m_t, m_b] \\
  &-& 8 (C_0 + C_{11}) [p_2 - p_3, -p_2, m_b, m_t, m_b] \\
  &+& 8 (m_{H^{\pm}}^2 (D_{11} - D_{12} + 2 D_{25} - 2 D_{26}) +
       m_t^2 (D_0 + D_{13}) - m_b^2 D_{13} - \\
  & &  2 (p_1 \cdot p_2) (D_{13} + D_{23}) +
       2 (p_1 \cdot p_3) (D_{13} + D_{26}) )
      [p_1, -p_1 + p_3, -p_2, m_t, m_b, m_t, m_b] \\
  &+& 8 (m_{H^{\pm}}^2 (D_{11} - D_{12} + 2 D_{13} + 2 D_{25}) +
      m_t^2 (D_0 + D_{13}) - m_b^2 D_{13} + \\
  & &  2 (p_1 \cdot p_2) (D_{26} - D_{23}) -
       2 (p_1 \cdot p_3) (D_{13} + D_{26}) )
      [p_1, p_2 - p_3, -p_2, m_t, m_b, m_t, m_b] ).
\end{array}
\eqno{(A-42)}$$

$$ f_9=f_{9a}+f_{9b}+f_{9c},\eqno{(A-43)}$$
$$\begin{array}{lll}
f_{9a} &=&
   16 g_{-} g_{+} (
      (D_{12} - D_{11} - D_{13} - D_{21} + D_{24} - D_{25})
      [-p_1, p_1 - p_3, -p_4, m_t, m_b, m_t, m_t] \\
  &+& (D_{12} - D_{11} - D_{13} - D_{21} + D_{24} - D_{25})
       [-p_1, -p_2 + p_3, -p_3, m_t, m_b, m_t, m_t]),
 \end{array}
\eqno{(A-44)}$$

$$ f_{9b}=f_{9a}~~(m_{b}\leftrightarrow m_{t}), \eqno{(A-45)}$$
$$\begin{array}{lll}
f_{9c} &=&
     32 i g_s^2 g_{Z H^{\pm}H^{\mp}} g_{Ztt}^{a} A_{Z} \cdot
       (C_{23} - C_{22})[p_3,-p_1-p_2,m_t,m_t,m_t]\\
  &+ & 16 g_{-} g_{+} (
   (D_{22} - D_{24} + D_{25} - D_{26})
       [p_1, -p_1 + p_3, -p_2, m_t, m_b, m_t, m_b] \\
  &+& (D_{13} + D_{25}) [p_1, p_2 - p_3, -p_2, m_t, m_b, m_t, m_b]).
 \end{array}
\eqno{(A-46)} $$

$$ f_{10}=f_{10a}+f_{10b}+f_{10c},\eqno{(A-47)}$$
$$\begin{array}{lll}
f_{10a} &=&
   16 g_{-} g_{+} (
         (D_{11} - D_{12} + D_{13} + D_{21} - D_{24} + D_{25})
         [-p_1, p_1 - p_3, -p_4, m_t, m_b, m_t, m_t] \\
  &+& (D_{11} - D_{12} + D_{13} + D_{21} - D_{24} + D_{25})
          [-p_1, -p_2 + p_3, -p_3, m_t, m_b, m_t, m_t] ),

\end{array}\eqno{(A-48)}$$

$$ f_{10b}=f_{10a}~~(m_{b}\leftrightarrow m_{t}), \eqno{(A-49)}$$
$$\begin{array}{lll}
f_{10c} &=&
     16 i g_s^2 g_{Z H^{\pm}H^{\mp}} g_{Ztt}^{a} A_{Z} \cdot
   (C_{11} + C_{21} +2 C_{22} - 2 C_{23})
       [p_4,-p_1-p_2,m_t,m_t,m_t]   \\
  &+ & 16 g_{-} g_{+} (
 - (D_{13} + D_{25})[p_1, -p_1 + p_3, -p_2, m_t, m_b, m_t, m_b] \\
  &+& (D_{24} - D_{22} - D_{25} + D_{26})
      [p_1, p_2 - p_3, -p_2, m_t, m_b, m_t, m_b] ).
 \end{array}
\eqno{(A-50)}$$

$$ f_{11}=f_{11a}+f_{11b}+f_{11c},\eqno{(A-51)}$$
$$\begin{array}{lll}
f_{11a} &=&
   16 g_{-} g_{+} (
         (D_{22} - D_{13} - D_{24} - D_{26})
       [-p_1, -p_2 + p_3, -p_3, m_t, m_b, m_t, m_t] \\
   &-& (D_{13} + D_{25}) [-p_1, p_1 - p_3, -p_4, m_t, m_b, m_t, m_t] ),
 \end{array}
\eqno{(A-52)}$$

$$ f_{11b}=f_{11a}~~(m_{b}\leftrightarrow m_{t}), \eqno{(A-53)}$$
$$\begin{array}{lll}
f_{11c} &=&
  32 i g_s^2 g_{Z H^{\pm}H^{\mp}} g_{Ztt}^{a} A_{Z} \cdot
      (C_{23} - C_{22})[p_3,-p_1-p_2,m_t,m_t,m_t]  \\
   &+ & 16 g_{-} g_{+} (
    (D_{26} - D_{23}) [p_1, -p_1 + p_3, -p_2, m_t, m_b, m_t, m_b] \\
   &+& (D_{26} - D_{23}) [p_1, p_2 - p_3, -p_2, m_t, m_b, m_t, m_b] ).
 \end{array}
\eqno{(A-54)}$$

$$ f_{12}=f_{12a}+f_{12b}+f_{12c},\eqno{(A-55)}$$
$$\begin{array}{lll}
f_{12a} &=&
    16 g_{-} g_{+} (
         (D_{13} - D_{22} + D_{24} + D_{26})
         [-p_1, p_1 - p_3, -p_4, m_t, m_b, m_t, m_t] \\
   &+& (D_{13} + D_{25}) [-p_1, -p_2 + p_3, -p_3, m_t, m_b, m_t, m_t] ),
 \end{array}
\eqno{(A-56)}$$

$$ f_{12b}=f_{12a}~~(m_{b}\leftrightarrow m_{t}),\eqno{(A-57)} $$
$$\begin{array}{lll}
f_{12c} &=&
    16 i g_s^2 g_{Z H^{\pm}H^{\mp}} g_{Ztt}^{a} A_{Z} \cdot
    (C_{11}  + C_{21} +2 C_{22} - 2 C_{23})
      [p_4,-p_1-p_2,m_t,m_t,m_t]  \\
   &+ & 16 g_{-} g_{+} (
    (D_{23} - D_{26}) [p_1, -p_1 + p_3, -p_2, m_t, m_b, m_t, m_b] \\
   &+& (D_{23} - D_{26}) [p_1, p_2 - p_3, -p_2, m_t, m_b, m_t, m_b] ).
 \end{array}
\eqno{(A-58)}$$

In the above form factor expressions we used the following definitions.
$$ g_{\gamma tt} = \frac{-2 i}{3} e,                              $$
$$ g_{Ztt}^v = \frac{-i g}{4 \cos \theta}(1-\frac{8}{3} \sin^2 \theta),   $$
$$ g_{Ztt}^a = \frac{i g}{4 \cos \theta},                         $$
$$ g_{\gamma H^{\pm}H^{\mp}} = -i e,                              $$
$$ g_{ZH^{\pm}H^{\mp}} = \frac{-i g \cos 2\theta}{2 \cos \theta}, $$
$$ g_{H^0tt} = -i g \frac{m_t \sin \alpha}{2 m_W \sin \beta}, $$
$$ g_{h^0tt} = -i g \frac{m_t \cos \alpha}{2 m_W \sin \beta}, $$
$$ g_{H^0 H^{\pm} H^{\mp}} = -i g (m_W \cos(\beta - \alpha) -
m_Z \frac{\cos 2 \beta}{2 \cos \theta_{W}} \cos(\beta + \alpha)), $$
$$ g_{h^0  H^{\pm} H^{\mp}} = -i g (m_W \sin(\beta - \alpha) +
   m_Z \frac{\cos 2 \beta}{2 \cos \theta_{W}} \sin(\beta + \alpha)), $$
$$ g_{+} = \frac{i g}{2 \sqrt{2} m_W} (m_b \tan \beta + m_t \cot \beta), $$
$$ g_{-} = \frac{i g}{2 \sqrt{2} m_W} (m_b \tan \beta - m_t \cot \beta), $$
$$ A_{\gamma} = \frac{-i}{s},       $$
$$ A_{Z} = \frac{-i}{s - m_{Z}^2 }, $$
$$ A_{H^0} = \frac{i}{s - m_{H^0}^2 + i \Gamma_{H^0} m_{H^0}}, $$
$$ A_{h^0} = \frac{i}{s - m_{h^0}^2 + i \Gamma_{h^0} m_{h^0}}. \eqno{(A-59)} $$

The arguments of two-point, three-point and four-point integral functions
are written at the end of formulae in paranthesis. The repalcement
$m_b \rightarrow - m_b$ is not performed on the arguments of loop
intergral functions B, C and D.

\vskip 20mm
{\Large{\bf Figure captions}}
\vskip 5mm
\noindent
{\bf Fig.1} The Feynman diagram of subprocess $gg \rightarrow H^{+} H^{-}$.
(a,b,c,d)s-channel diagrams. (e $\sim$ j)box diagrams including
t- and u-channels.
\vskip 3mm
\noindent
{\bf Fig.2(1)}Total cross sections of the subprocess $gg \rightarrow
H^{+} H^{-}$ as the functions of the $\sqrt{\hat{s}}$ with $m_{h^{0}}=100~GeV$,
$m_{H^{0}}=150~GeV$ and $m_{H^{\pm}}=150~GeV$. The full line is for
$\tan{\beta}=1.5$. The dashed line is for $\tan{\beta}=30$.
\vskip 3mm
\noindent
{\bf Fig.2(2)}Total cross sections of the subprocess $gg \rightarrow
H^{+} H^{-}$ as the functions of the $\sqrt{\hat{s}}$ with $m_{h^{0}}=100~GeV$,
$m_{H^{0}}=150~GeV$ and $m_{H^{\pm}}=300~GeV$. The full line is for
$\tan{\beta}=1.5$. The dashed line is for $\tan{\beta}=30$.
\vskip 3mm
\noindent
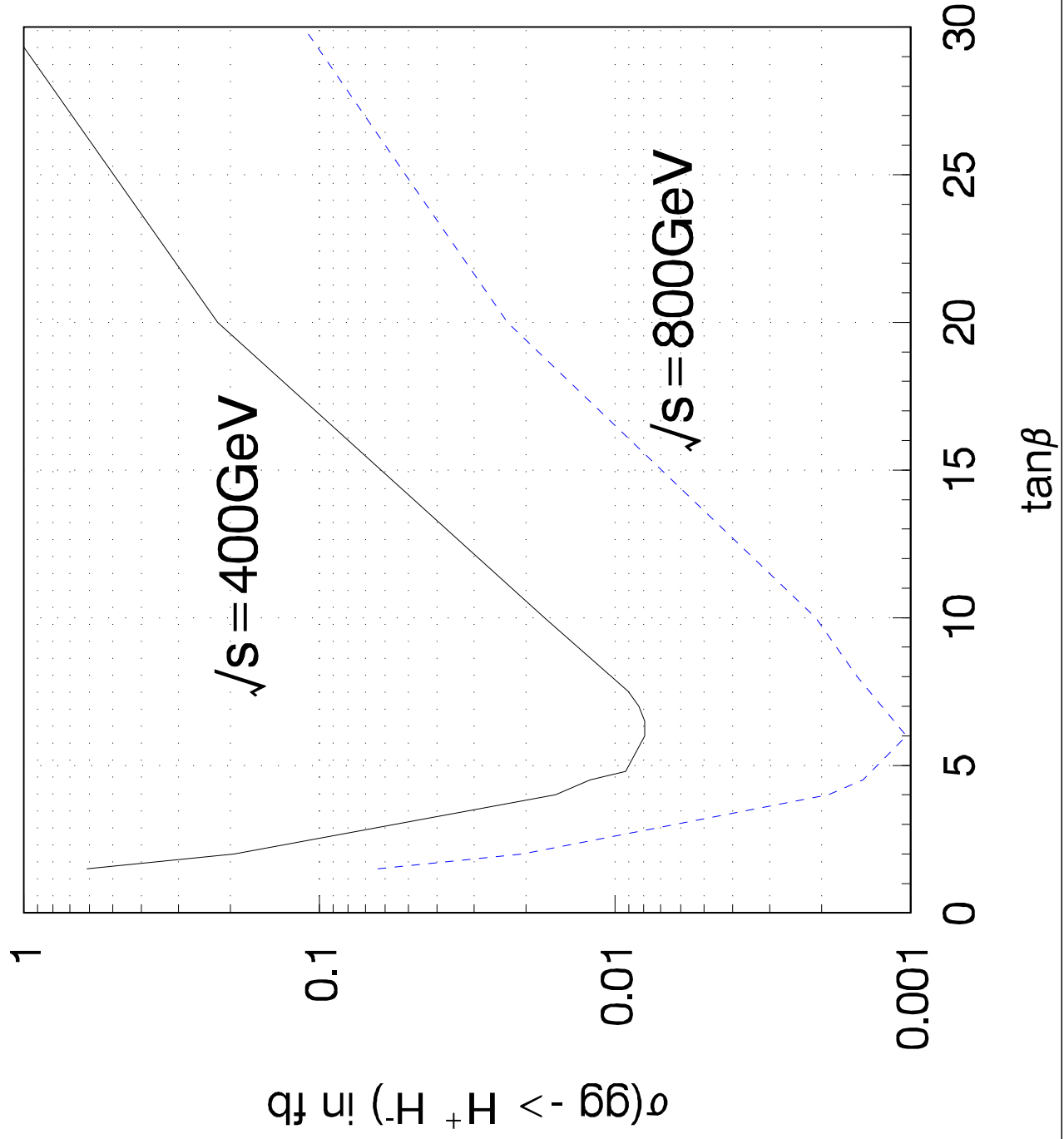
{\bf Fig.3}Total cross sections of the subprocess $gg \rightarrow
H^{+} H^{-}$ as the functions of the $\tan{\beta}$ with $m_{h^{0}}=100~GeV$,
$m_{H^{0}}=150~GeV$ and $m_{H^{\pm}}=150~GeV$. The full line and dashed line
correspond to $\sqrt{\hat{s}}=400~GeV$ and $\sqrt{\hat{s}}=800~GeV$
respectively.
\vskip 3mm
\noindent
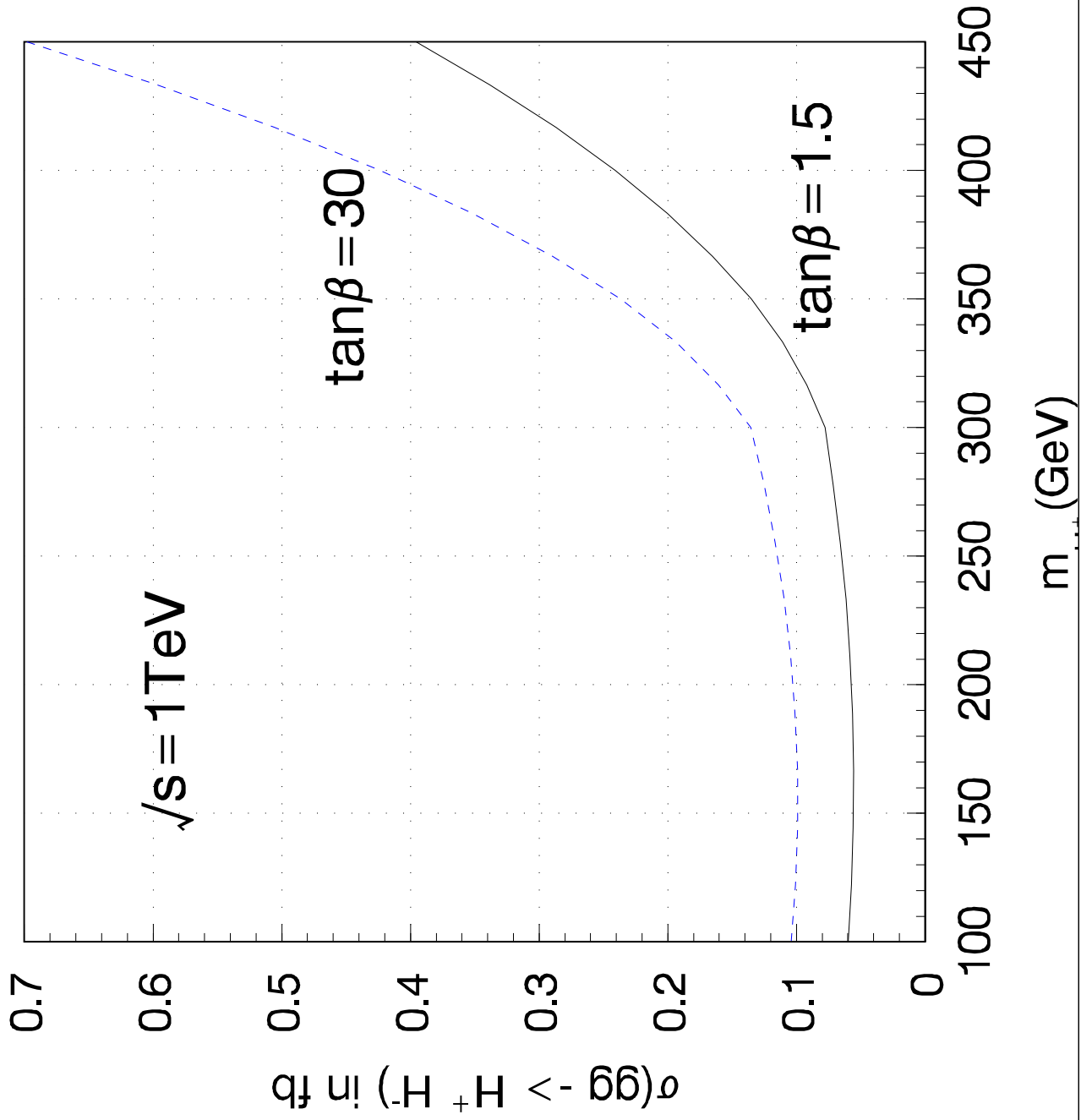
{\bf Fig.4}Total cross sections of the subprocess $gg \rightarrow
H^{+} H^{-}$ as the functions of the $m_{H^{\pm}}$ with $m_{h^{0}}=100~GeV$,
$m_{H^{0}}=150~GeV$ and $\sqrt{\hat{s}}=1~TeV$. The full line is for
$\tan{\beta}=1.5$. The dashed line is for $\tan{\beta}=30$.
\vskip 3mm
\noindent
{\bf Fig.5(1)}Total cross sections of the process $pp \rightarrow gg
\rightarrow H^{+} H^{-} + X$ as the functions of the $\sqrt{s}$ with
$m_{h^{0}}=100~GeV$, $m_{H^{0}}=150~GeV$ and $m_{H^{\pm}}=150~GeV$.
The full line is for $\tan{\beta}=1.5$. The dashed line is for
$\tan{\beta}=30$.
\vskip 3mm
\noindent
{\bf Fig.5(2)}Total cross sections of the process $pp \rightarrow gg
\rightarrow H^{+} H^{-} + X$ as the functions of the $\sqrt{s}$ with
$m_{h^{0}}=100~GeV$, $m_{H^{0}}=150~GeV$ and $m_{H^{\pm}}=300~GeV$.
The full line is for $\tan{\beta}=1.5$. The dashed line is for
$\tan{\beta}=30$.

\vskip 3mm
\end{large}
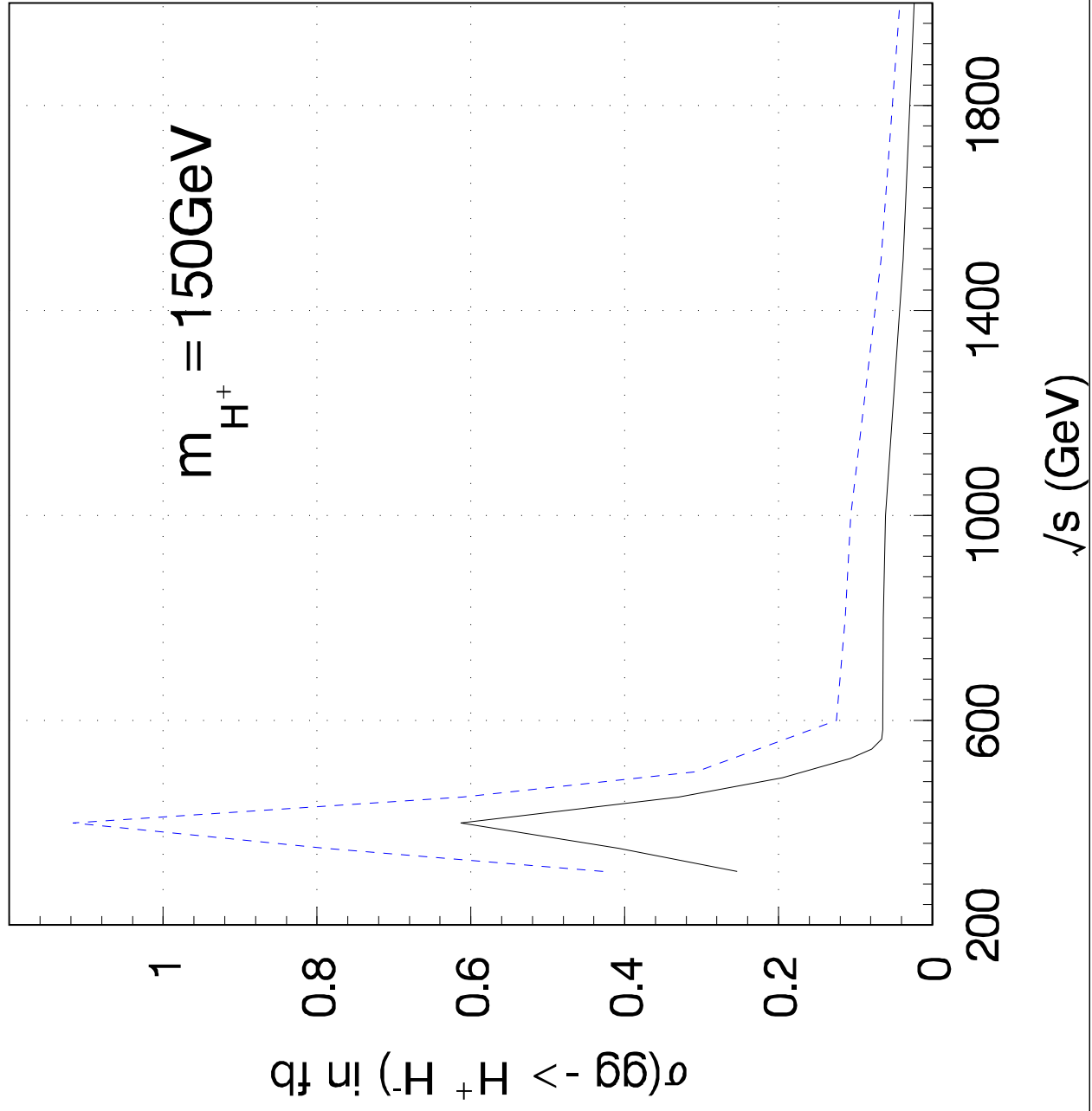
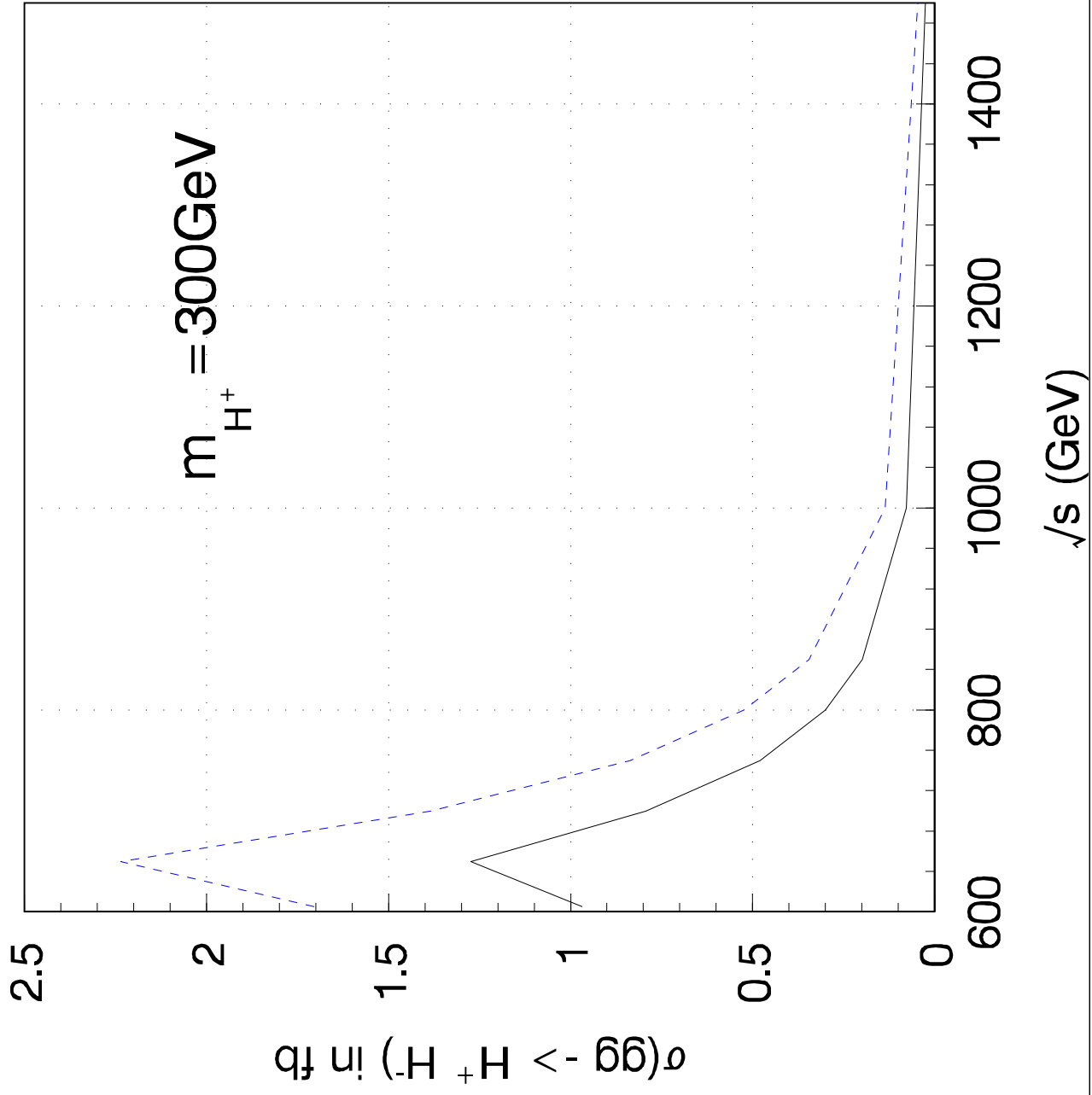
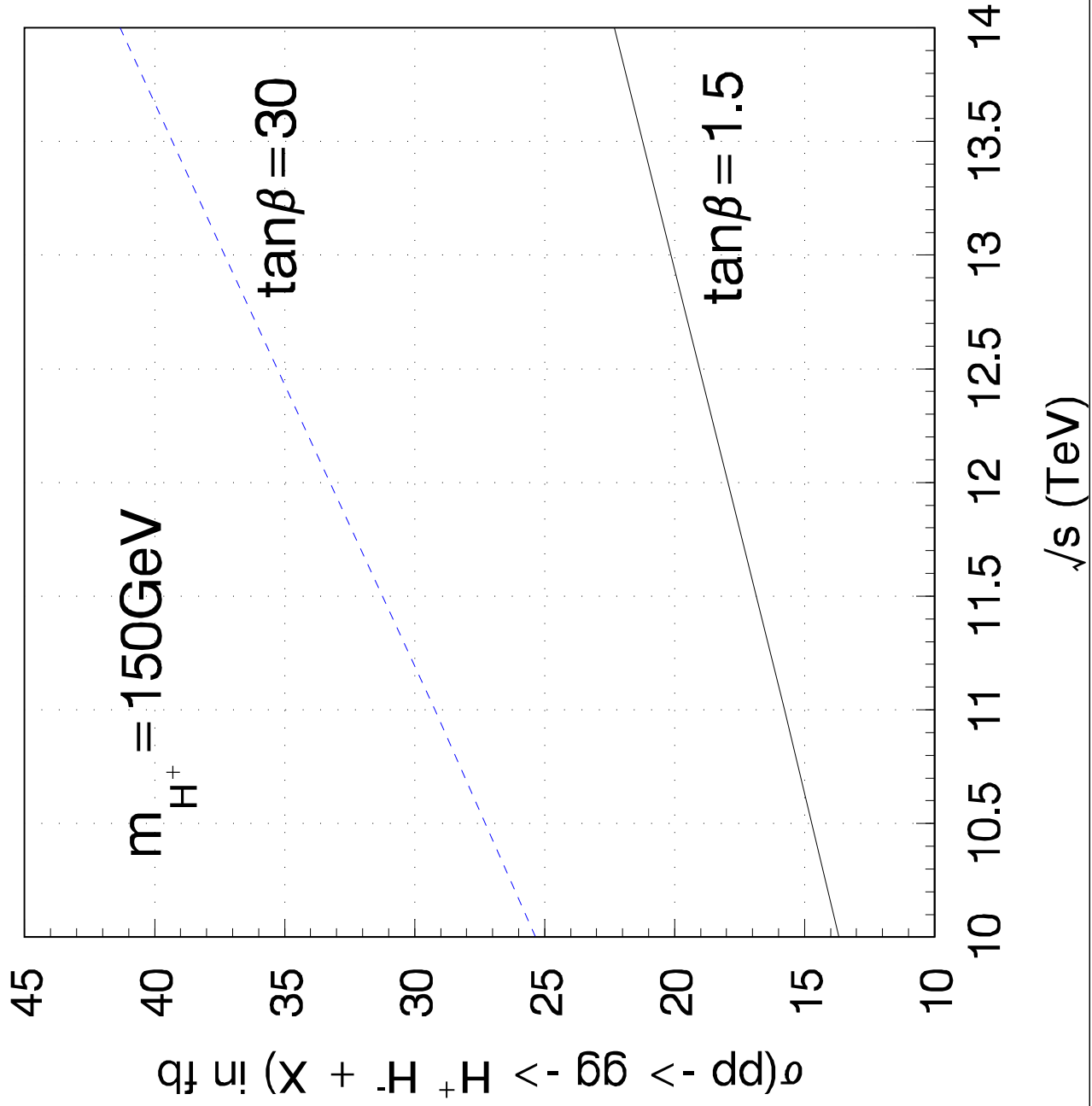
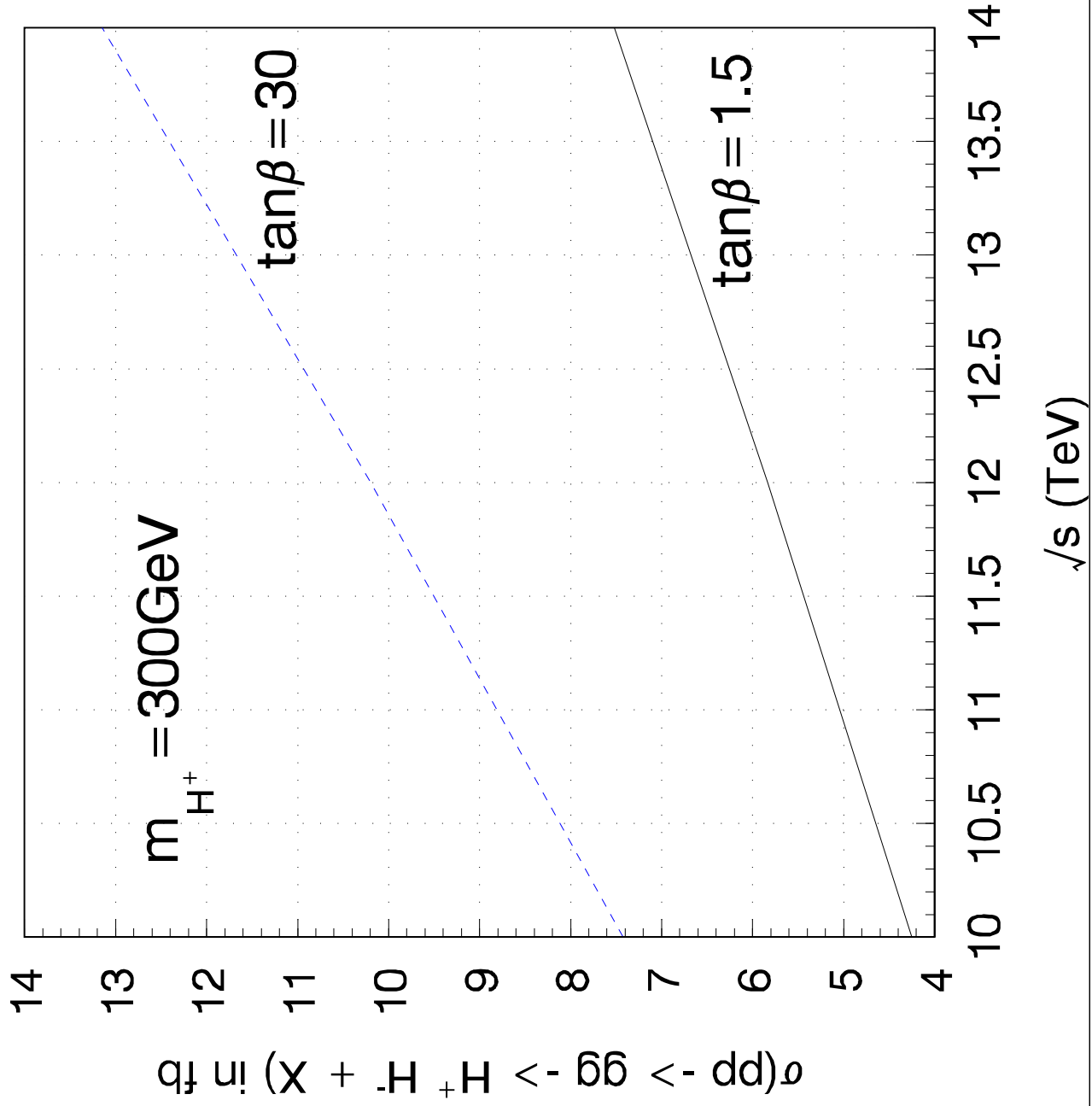
\end{document}